\documentclass[12pt]{article}
\usepackage{epsf}

\newcommand{\be}{\begin{equation}}
\newcommand{\ee}{\end{equation}}
\newcommand{\bea}{\begin{eqnarray}}
\newcommand{\eea}{\end{eqnarray}}
\newcommand{\vap}{\vartheta}

\newcommand{\parp}{\partial}
\newcommand{\vet}{\overline}
\title{ \hfill $\mbox{\small{
$\stackrel{\rm\textstyle SU{-}4240{-}709\quad}
{\rm \textstyle \quad}
$}}$ \\[1truecm]
The Geometrical Structure of 2D Bond-Orientational Order}

\author{\\\\ Mark Bowick\thanks{\tt bowick@physics.syr.edu} \,
 and Alex Travesset \thanks{\tt alex@suhep.phy.syr.edu}
\\\\ 
Physics Department, Syracuse University,\\
Syracuse, NY 13244-1130, USA }

\date{}

\begin{document}

\begin{titlepage}
\maketitle
\begin{abstract}
We study the formulation of bond-orientational 
order in an arbitrary two dimensional geometry. 
We find that bond-orientational order is properly formulated 
within the framework of differential geometry with torsion. 
The torsion reflects the intrinsic frustration for two-dimensional 
crystals with arbitrary geometry. 
Within a Debye-Huckel approximation, torsion may be identified as 
the density of dislocations. Changes in the geometry of the system 
cause a reorganization of the torsion density that preserves
bond-orientational order. 
As a byproduct, we are able to derive several identities involving
the topology, defect density and geometric invariants such as Gaussian
curvature. The formalism is used to derive the general free energy for 
a 2D sample of arbitrary geometry, both in the crystalline and
hexatic phases. Applications to conical and spherical geometries are 
briefly addressed.

\end{abstract}
\end{titlepage}

\section{Introduction}\label{sect__introd}

According to the KTNHY theory \cite{NEL1,YOU1,NEL2},
two-dimensional melting in the plane is a two stage defect-driven 
process involving continuous crystalline-to-hexatic and
hexatic-to-fluid transitions. The intermediate hexatic phase is
characterized by quasi-long-range bond orientational order and
positional disorder. In a wide variety of settings one may expect
to encounter geometries more general than the plane. In the theory of
membranes the geometry itself fluctuates. External forces may act to
bend the geometry to some fixed curved surface. This raises the
challenging problem of generalizing the established KTNHY theory to 
substrates with some arbitrary geometry. In this paper we discuss a 
complete geometrical formulation of bond-orientational order
\cite{Str} (hexatic and
crystalline) that provides a framework in which
the interaction of defects, geometry and topology are easily
formulated. 

To begin, consider a hypothetical flat monolayer displaying 
bond-orien\-tat\-ion\-al order at zero temperature. Bonds may be represented 
as six vectors 
at each point of the plane, with a relative angle of $\frac{\pi}{3}$. 
The six vectors at any given point of the plane are {\em parallel}
to those at any other point.
In other words, if we know the vectors at some point
we can reconstruct the vectors at any other point by parallelism.
Imagine now adiabatically deforming the plane of the monolayer 
to an arbitrary curved surface. Intuitively we would expect the
bond-orientational order to be stable to this deformation. The curvature of the 
monolayer, however, now implies an associated Gaussian curvature. 
As well known from differential geometry there is no intrinsic notion
of parallelism on curved surfaces. To define parallel transport we
require a rule specified via a connection. In general parallel
transport of a vector between two points will depend on the path. 
For our problem this means that the bond angle at an arbitrary point
on the surface has no path-independent meaning. For the standard choice 
of connection (the Levi-Civita connection) the bond angle at some
point reached by two distinct paths from a reference point will depend
on the total Gaussian curvature enclosed by the paths. 
This sensitive dependence on the path suggests that bond-orientational
order is incompatible with any curved geometry.

\begin{figure}[htb]
\epsfxsize=5 in \centerline{\epsfbox{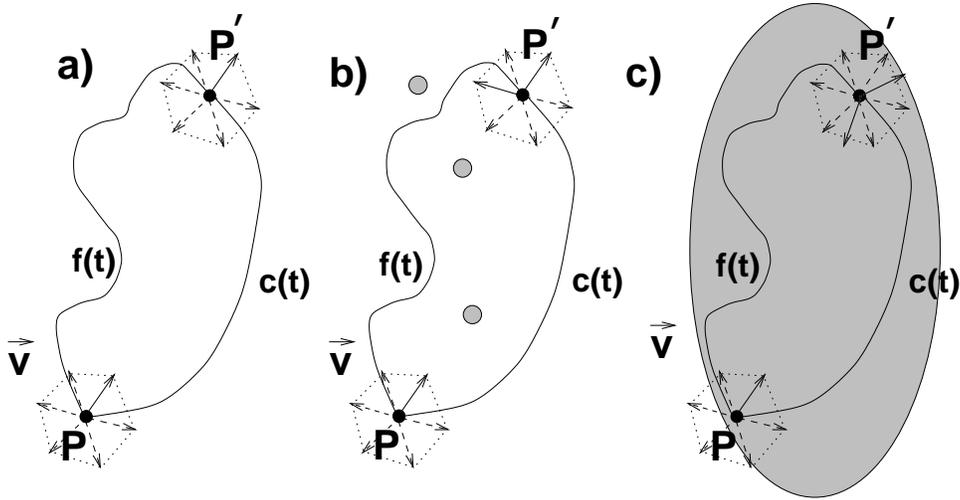}}
\caption{The result of parallel transport of ${\bf V}$ from $P$ to $P'$ 
a) phase with
no disclinations: the result is path-independent 
b) phase with a few isolated disclinations, marked as
grey spots: the result may differ by an angle of $\frac{\pi}{3}$ but
bond-orientational order is preserved  
c) isotropic fluid phase: parallel transport is
completely ambiguous.}
\label{fig__disclination}
\end{figure}

This mathematical argument, however, contradicts our intuition
that bond-orientational order should be stable to deformation of the surface.
This suggests that a more physical connection exists for defining
bond-orientational order on curved surfaces. To see this note that  
bond-orien\-tational order very generally implies the existence of a
frame of six {\em bonds} forming a $\frac{\pi}{3}$ angle at each point 
of the sample. 
Given this data we can define a {\em path-independent} parallel
transport in the following way. Take a vector ${\bf V}$ at point $P$.
This forms a certain angle with the frame. Now we can transport ${\bf
V}$ to the point $P'$ by demanding that ${\bf V}$ forms the 
same angle to the given frame at $P'$. By construction this {\em
physical} parallel transport does not depend on the path and is
therefore unambiguous \footnote{The fact that this parallel transport is 
unambiguous can be made more explicitly if, before starting the experiment,
we mark all the bonds pointing, say, positively in the $x$-direction,
which can be done unambiguously in the absence of free disclinations.}.

Clearly the connection that corresponds to the parallel transport 
defined above must be different from the standard Levi-Civita
connection. We will argue below that the mathematical realization of
this physical connection appropriate for defining bond-orientational
order involves a new degree of freedom, a geometrical quantity called
torsion. The difference angle between two parallel transported vectors 
will depend on the enclosed curvature of the new connection with
torsion. This connection may be tuned to be curvature-free so that
parallel transport is path-independent, as desired. 

Although torsion may appear at this stage as merely a mathematical
trick, a number of authors \cite{TOR} have identified torsion as
necessary for describing a system with a high density 
of {\em dislocations}, such as in the hexatic phase. This is certainly the 
case if the density of dislocations is well approximated by a continuous 
density, corresponding to the Debye-Huckel approximation. 
We will see more generally, however, that torsion accounts for the intrinsic
frustration of a 2d configuration on a curved surface.

This curvature of the connection with torsion also has a clear physical 
interpretation as being proportional to the density of {\em
disclinations}. Raising the temperature slightly will excite
disclinations. Although the zero curvature condition is violated, it
is such that  parallel transport is ambiguous only up to an angle $\pi/3$, and the hexatic
order is preserved. As the temperature is raised further the growing
density of free disclinations will eventually melt the hexatic to a
fluid. The high density of disclinations in this phase may be
represented by a continuous disclination density. In this case  the
parallel transported bond angle is completely ambiguous, which is
equivalent to saying that bond angles are meaningless in the isotropic
fluid phase. 

The organization of the paper is as follows: in the 
next section we introduce the necessary background 
in differential geometry, which will also serve to fix our notation.
In section~\ref{SECT__bondor} we provide the physical interpretation
of torsion and derive some general relations involving the geometry,
the defect density and the torsion. In section~\ref{SECT__hex} we 
reformulate the case of a simple flat monolayer in terms of the new 
formalism: this is easily generalized to an
arbitrary geometry. We end the paper with some conclusions. In 
the appendix we treat the example of a cone displaying hexatic order 
as an explicit application of our formalism.

\section{Differential geometry background}\label{sect__diff_geom}

In this section we briefly emphasize those concepts in differential 
geometry that play a major role in the subsequent analysis. For further
details we refer the reader to the literature \cite{DIFF}.

\subsection{Basic differential geometry}

A $D$-dimensional surface embedded in $d$-dimensional Euclidean space is 
described by $d${-}dimensional functions 
$R^{\alpha}({\bf x}) \ , \alpha=1,\cdots,d$ of $D$  
coordinates $x^{\mu} \ , \mu=1, \cdots D$, with $D=2$ being the physically
interesting case.

The fact that the surface is embedded in ${\bf R}^d$ provides a 
natural metric,
\be\label{met_surface}
ds^2=g_{\mu \nu} dx^{\mu} \otimes dx^{\nu} \ , \ 
g_{\mu \nu}=\parp_{\mu}{\vec R} \cdot\parp_{\nu} {\vec R} \ .
\ee
Introducing the vielbeins, ${e^a}_{\mu}$, it is possible to rewrite
the previous equation as
\be\label{diag_surf}
ds^2=\delta_{a b} \vap^a \otimes \vap^b \ ,
\ee
where $\vap^{a}={e^a}_{\mu} dx^{\mu}$. It is easy to check that
the tangent vectors 
\be\label{def_tangent}
e_a={e^{\mu}}_a \frac{\parp}{\parp x^{\mu}} \ , \ a=1,\cdots,D \ ,
\ee
where ${e^{\mu}}_a {e^a}_{\nu}=g_{\mu \nu}$,
define a basis of orthonormal vectors. The dot product of any
two vectors ${\bf u}=u^a e_a$ and ${\bf v}=v^a e_a$ in tangent space 
is ${\bf u} \cdot {\bf v}=\delta_{ab}u^a v^b$, as if the metric
were flat. This formalism is called the non-coordinate basis.
Hereafter we use the first letters of the Latin alphabet when we
work in a non-coordinate basis, and Greek indices for the 
coordinate basis.

The space of all $r$-dimensional forms of a manifold $M$
$\Omega^{r}(M)$ plays an important role. The metric allows one to define a 
natural isomorphism, the Hodge-star operator
\be\label{iso_forms}
\Omega^{r}(M) \stackrel{\ast}{\rightarrow} \Omega^{D-r}(M) \ ,
\ee
defined as the linear map which in a non-coordinate acts as
\be\label{iso_form_precise}
\ast(\vap^{a_1}\wedge\vap^{a_2}\wedge\cdots\wedge\vap^{a_r})=
\frac{1}{(D-r)!}{\epsilon^{a_1\cdots a_r}}_{a_{r+1}\cdots a_D}
(\vap^{a_{r+1}}\wedge\cdots\wedge\vap^{a_D}) ,
\ee
where $\epsilon^{1 2 \cdots D}=1$. We may define an inner 
product in $\Omega^{r}(M)$. If 
$\omega=\frac{1}{r!}\omega_{a_1\cdots a_r}\vap^{a_1}\wedge\cdots\vap^{a_r}$
and
$\nu=\frac{1}{r!}\nu_{a_1\cdots a_r}\vap^{a_1}\wedge\cdots\vap^{a_r}$
\be\label{inner_prod}
(\omega|\nu)=\int_M \omega \wedge \ast \nu=\int_M \omega_{a_1\cdots a_r}
\nu^{a_1 \cdots a_r} \sqrt{|g|} dx^1\cdots dx^D \ ,
\ee
the integral being over the whole manifold $M$.

The adjoint $d^{\dagger}$ of the exterior derivative is defined as
the linear operator satisfying
\be\label{d_dagger}
(d \alpha|\beta)=(\alpha|d^{\dagger} \beta) \ .
\ee
It is an easy exercise to check that $d^{\dagger}=(-1)^{Dr+D+1} \ast d \ast$.
The Laplacian is the operator
\be\label{def_laplacian}
\Delta=d d^{\dagger}+d^{\dagger} d \ .
\ee
Let us recall that the Laplacian is independent of the connection.

\subsection{Connections}

A connection on a manifold $M$ specifies how tensors are transported along
a curve. It allows one to define a covariant derivative $\nabla_{\mu}$ on
a vector field $U$,
\be\label{cov_der}
\nabla_{\mu} U^{\nu}=\parp_{\mu}U^{\nu}+\Gamma^{\nu}_{\mu \sigma} U^{\sigma} 
\ ,
\ee
where $\Gamma^{\nu}_{\mu \sigma}$ are the connection coefficients.
The parallel transport of a vector $V^{\mu}$ along a curve $c^{\mu}$ is 
fixed by demanding that is be covariantly constant: 
\be\label{def_par_transp}
\nabla_c V^{\mu}=\frac{d c^{\nu}(t)}{d t} \nabla_{\nu} V^{\mu}=0 \ .
\ee
It is convenient to define the connection coefficients in a non-coordinate 
basis:
\be\label{cov_noncoord}
\nabla_a e_{b}={\Gamma^{c}}_{a b} e_c \ ,
\ee
and a connection one-form via
\be\label{con_oneform}
{\omega^{a}}_{b}={\Gamma^a}_{cb}\vap^c \ .
\ee
We consider only metric connections that preserve the norm of a vector
under parallel transport, a property called metric compatibility.
This property constrains the connection one-form to satisfy
\be\label{metric_comp}
{\omega^a}_b=-{\omega^b}_a \ ,
\ee
which in $D=2$ further simplifies to
${\omega^a}_{b}=-{\epsilon^a}_{b} \Omega$,
$\Omega=\Omega_{a} \vap^a$.
Connection coefficients do not transform as a tensor but it 
is possible to construct two tensorial objects out of them, the
torsion and the curvature. For the sake of completeness, we 
write them acting on vector fields $X,Y$,
\bea
T(X,Y)&=&\nabla_X Y -\nabla_Y X - [ X, Y ]
\nonumber\\
R(X,Y)&=&\nabla_X \nabla_Y -\nabla_Y \nabla_X  - \nabla_{ [ X, Y ] }
\eea
Expressions turn out simpler in a non-coordinate
basis. Defining torsion and curvature 2-forms by
\bea\label{def_two_forms}
T^{a}&=&\frac{1}{2} {T^a}_{bc} \vap^b \wedge \vap^c
\nonumber\\
{R^a}_b&=&\frac{1}{2} {R^a}_{bcd} \vap^c \wedge \vap^d \ ,
\eea
they are related to the connection one-form by 
Cartan's structure equations, which read for the case $D=2$,
\bea\label{cartan_struc}
T^a&=&d\vap^{a}-\Omega \wedge \vap^{b} {\epsilon^{a}}_{b} \
\\\nonumber
\frac{1}{2}{\epsilon_a}^b {R^a}_b&=&d\Omega \ .
\eea
Recall that the curvature 2-form is locally an exact form. 
The expressions in the coordinate basis follow easily from those 
in the non-coordinate basis from the tensorial nature of curvature
and torsion, namely
${T^{\mu}}_{\nu \rho}={e^{\mu}}_a {e^b}_{\nu} {e^c}_{\rho} {T^a}_{bc}$ and
${R^{\mu}}_{\nu \rho \sigma}={e^{\mu}}_a {e^b}_{\nu} {e^c}_{\rho} 
{e^d}_{\sigma} {R^a}_{bcd}$. 

\subsection{Some global considerations}

The global structure of the manifold places important constraints on
the differential geometry of the surface. From Eq.~\ref{cartan_struc}
the curvature 2-form is locally an exact form, but that is not the case
globally. For a manifold without boundaries, integrating over the whole 
manifold we have
\be\label{Gauss_Bonnet}
\int_M \frac{1}{2}{\epsilon_a}^b {R^a}_b= 2 \pi \sum_{\alpha}Ind_{\alpha} = 
2 \pi \chi \ ,
\ee
where $Ind$ are the indices of the vector field associated with $\Omega$ 
and $\chi$ is the Euler characteristic of the manifold. If $R$ is the 
Riemannian curvature this is the usual Gauss-Bonnet theorem.

\section{The geometry of bond-orientational order}\label{SECT__bondor}

We now consider a two-dimensional sample exhibiting bond-orientational
order and forming an arbitrary shape in d-dimensional space, specified
by the embedding $R^{\alpha}({\bf x})$. Within the sample, distances 
are measured according to the metric Eq.~\ref{met_surface}. 

\begin{figure}[htb]
\epsfxsize=3 in \centerline{\epsfbox{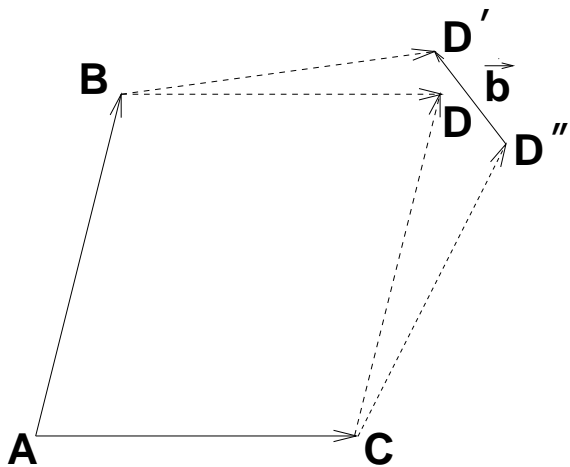}}
\caption{Parallel Transport in the presence/absence of Torsion.}
\label{fig__torsion}
\end{figure}

The main feature of connections with non-vanishing torsion
(non-sym\-metric connections) is that infinitesimal 
reference vectors fail to close when parallel transported along each
other, as illustrated in  Fig.~\ref{fig__torsion}.
If a crystal is forced to lie on an arbitrary geometry, 
the atoms in the crystal cannot be all equally separated by a distance
$a$. This implies that parallel transport along the bonds joining the
nearest-neighbors will fail to close, as shown in Fig.~\ref{fig__torsion}. 
Thus we see that torsion is related to the intrinsic frustration
experienced by a 2D crystal on an arbitrary surface.

In the hexatic phase there is a similar interpretation of torsion,
except that there will be a large number of
dislocations. This is specified by a Burgers vector density
\be\label{dens_disl}
{\bf b}({\bf x})=
\frac{1}{\sqrt{g}({\bf x})}\sum_i {\bf b}_i^L \delta({\bf x}_i-{\bf x})
\ee
where ${\bf b}_i^L$ is the microscopic Burgers vectors with origin at 
point ${\bf x}_i$. In a highly dislocated medium, the discreteness of
the Burgers vector density ${\bf b}(x)$ may be approximated 
by a continuous vector density, the so-called Debye-Huckel 
approximation.

The density ${\bf b}(x)$ is, in fact, directly related to the
torsion invariant appearing in Eq.~\ref{def_two_forms},
as noted by several authors \cite{TOR}. 
The torsion two-form is
\be\label{act_torsion}
T^a=v^a \vap^1\wedge\vap^2=v^a \Omega_M \ ,
\ee
where $\Omega_M=\sqrt{|g|({\bf x})} dx^1\wedge dx^2$ is the volume form. 
Since $v^a$ is a vector it can be written in a coordinate 
basis $v^{\mu}={e^{\mu}}_a v^a$. Its geometric interpretation is
illustrated in Fig.~\ref{fig__torsion}. Let us assume non-zero torsion 
at point $A$. Consider the vector $\vet{AC}$ and parallel transport 
it along vector $\vet{AB}$, the resulting vector is $\vet{BD^{\prime}}$.
If we parallel transport vector $\vet{AB}$ along $\vet{AC}$ we get
$\vet{CD''}$. Using the parallel transport Eq.~\ref{def_par_transp}
and Eq.~\ref{cov_der}, we may write to lowest order
\be\label{int_torsion}
\vet{D''D'}^{\mu}=\left\{ \Gamma^{\mu}_{\sigma \nu}-
\Gamma^{\mu}_{\nu \sigma} \right\}\vet{AB}^{\sigma}\vet{AC}^{\nu}
=v^{\mu}\times\mbox{Area of the region ($ABD'D''C)$} .
\ee
If ${\bf v}=0$ then $D'=D''=D$ and the parallelogram closes.

It is clear from the previous geometrical argument that torsion is
related to the continuous density of Burgers vector by
\be\label{tor_burg}
{\bf b}({\bf x})={\bf v}({\bf x}) \ ,
\ee
or $T^a=b^a \Omega_M$, in a non-coordinate basis. A highly dislocated
media physically is represented geometrically by a surface endowed
with a connection with torsion \cite{TOR}.

Clearly then  torsion is a necessary ingredient
in a geometrical description of bond-orientational order, as
it allows a connection form from which the bond angle is constructed 
by parallel transport. This interpretation of torsion is completely
general. In the Debye-Huckel limit we have the additional
identification of the torsion with the dislocation density.

In the absence of free disclinations we 
pointed out that this result should be independent of the path chosen.
This in turn translates into the mathematical requirement of vanishing 
curvature 2-form
\be\label{perf_hexatic}
{R^a}_b=0 \ .
\ee
This is a local condition, not always possible to fulfill globally, as 
for example in the case of spherical topology.  Anyway, 
Eq.~\ref{perf_hexatic} is fulfilled everywhere for
the case of a sample having the topology of a plane.

In this regime, the Cartan structure equations Eq.~\ref{cartan_struc}
read
\bea\label{Cartan_perf}
T^a&=&d\vap^{a}-\Omega \wedge \vap^{b} {\epsilon^a}_b \ ,
\\\nonumber
0&=&d\Omega  \ , 
\eea
The last equation implies $\Omega=d \theta$, where $\theta({\bf x})$ is a 
0-form, a function. Let us interpret its physical meaning.
The equation of parallel transport Eq.~\ref{def_par_transp} reads
\be\label{hex_parallel}
\frac{d V^{a}}{dt}-\Omega_{\mu}\frac{dc^{\mu}}{dt} {\epsilon^a}_b V^b
=0 \ ,
\ee
where $c^{\mu}(t)$ parametrizes a curve in the surface joining the
points $P$ and $P'$.
\be\label{par_sol}
V_{\pm}(P')=V_{\pm}(P) e^{-i\int_c \Omega_{\mu} dx^{\mu}} \ ,
\ee
In the hexatic phase with no free disclinations, 
Eq.~\ref{Cartan_perf} implies
\be\label{par_hex}
V_{\pm}(P')=V_{\pm}(P) e^{-i\theta(P')} \ ,
\ee
where $V_{\pm}=V^1\pm i V^2$ and $\theta(P)=0$. This   
parallel transport id manifestly path independent and
$\theta$ clearly is the bond angle at point $P'$. Dragging $P'$ along the
entire surface and performing the parallel transport Eq.~\ref{par_hex} 
we can unambiguously construct the bond angle at any point from the 
knowledge of the bond angle at $P$. We have succeeded in giving a precise 
mathematical formulation of the heuristic zero temperature experiment 
performed in the introduction.

\subsection{Introduction of free disclinations}

The sample may also contain some free disclinations, either as a
result of thermal fluctuations or because
topological constraints force them to appear. In either case,
the parallel transport can only be defined modulo an angle of  
$\frac{\pi}{3}$. The system responds to the raising of the
temperature by generating curvature, but only via defects that preserve 
overall hexatic order, the disclinations. In this way, the
curvature is directly related to the density of disclinations \cite{TOR}.

From these arguments, it is clear that a finite density of 
$N$ free disclinations is represented as
\bea\label{disclin_curv}
\frac{1}{2}{\epsilon_a}^b {R^a}_b&=&s({\bf x}) \Omega_M
\nonumber\\
s({\bf x})&=&\frac{\pi}{3}\frac{1}{\sqrt{|g|}}
\sum_{j=1}^{N} q_j \delta({\bf x}-{\bf x}_j) \ ,
\eea
where $q_i$ may be $+1$ (a five-fold disclination) or  
$-1$ (a seven-fold disclination). Higher integer charges are also
possible although rarely need to be considered.
If the surface is closed, we can integrate the previous equation 
over the whole manifold. With the aid of the Gauss-Bonnet theorem 
Eq.~\ref{Gauss_Bonnet} we find
\be\label{Euler_disclination}
\sum_{j=1}^{N} q_j=6\chi .
\ee
This shows that, apart from the torus with $\chi=0$, 
closed manifolds always contain a certain number of free disclinations.

With free disclinations, the Cartan structure equations (Eq.~\ref{cartan_struc})
become 
\bea\label{Hex_curv}
T^a&=&d\vap^{a}-\Omega \wedge \vap^{b} {\epsilon^a}_b \ ,
\nonumber\\
d \Omega&=& s({\bf x}) \Omega_M \ .
\eea
Away from the actual location of the disclinations $\Omega=d \theta$, but
now it is generally impossible for $\theta$ to be defined as a continuous 
function. It is easy to check that a vectors parallel transported from 
point $P$ to $P'$ 
following two different paths ${\bf c}(t)$ and ${\bf f}(t)$ paths (see 
Fig.~\ref{fig__disclination}) differ by an angle $\psi$
\be\label{ambig_z6}
\psi=\frac{\pi}{3} \sum_j q_j  \ ,
\ee
where $j$ runs over all charges within the region enclosed by the two
paths. The ambiguity is a multiple of  
$\frac{\pi}{3}$ and as expected hexatic order is preserved. Mathematically 
we have a connected manifold with a $Z_6$ holonomy. Disclinations 
may be regarded as orbifolds. 

Generally, we will write
\be \label{actu_omega}
\Omega= d \theta+\Omega_{sing} \ ,
\ee
explicitly displaying the bond angle $\theta$ and a 
a singular part, obviously corresponding to a vortex contribution. 
This is nothing but the procedure of separating  a regular part and a 
singular part in the standard treatment of the XY model \cite{XYM}.

There is more information encoded in the Cartan structure equations  
(Eq.~\ref{cartan_struc}). First of all, the Levi-Civita connection
is obtained imposing the vanishing of torsion,
\be\label{Levi_civ}
d \vap^{a}- \Omega^L \wedge \vap^b {\epsilon^a}_b=0 \ ,
\ee
where $\Omega^L$ is the connection form of the Levi-Civita connection.
We have the important relation
\be\label{Gauss_curv}
d \Omega^L=\frac{1}{2}{\epsilon^a}_b {R_a^G}^b=K({\bf x})\Omega_M \ ,
\ee
where ${R_a^G}^b$ is the standard Riemann tensor two-form and $K({\bf x})$ 
is the Gaussian curvature of the surface. Using this relation in
Eq.~\ref{Hex_curv}
\be\label{Cartan_hex}
 \Omega^L \wedge \vap^b {\epsilon^a}_b-
 \Omega \wedge \vap^b {\epsilon^a}_b=T^a \ .
\ee
Introducing the Burgers form $b=b_a({\bf x}) \vartheta^a=
b_{\mu}({\bf x})d x^{\mu}$, together with Eq.~\ref{tor_burg},
this may be rewritten as
\be\label{Final_form}
b=-\Omega+\Omega^L \ .
\ee
This is a very important equation. It relates the bond angle, computed  
from the actual connection $\Omega$ as indicated in Eq.~\ref{par_sol},
to the distribution of dislocations and to the Gaussian curvature of 
Eq.~\ref{Gauss_curv}. We can make this statement more apparent by applying
the exterior derivative to Eq.~\ref{Final_form}, and using Eqs.~\ref{Hex_curv}
and ~\ref{Gauss_curv}: 
\bea\label{Gauss_form}
K({\bf x})&=&s({\bf x})+\frac{1}{\sqrt{|g|}}\epsilon^{\mu \nu}
\parp_{\mu}b_{\nu}({\bf x})
\\\label{fancy_Gauss}
\frac{3}{\pi}K({\bf x})&=&\frac{1}{\sqrt{|g|}}\sum_{j=1}^{N}q_j
\delta({\bf x}-{\bf x}_j)+\frac{3}{\pi}
\frac{1}{\sqrt{|g|}}\epsilon^{\mu \nu} \parp_{\mu}b_{\nu}({\bf x}) \ .
\eea
Note that the coordination number of an atom $n_i$, the number of 
nearest-neighbors, is given by
\be\label{coord_numb} 
6-n_i=q_i=\int_{\Sigma} du \sqrt{g}\frac{1}{\sqrt{|g|}}\sum_{j=1}^{N}q_j 
\delta({\bf x}-{\bf x}_j)=\frac{3}{\pi}\int_{\Sigma} d u \sqrt{g} K
-\frac{3}{\pi}\int_{\partial \Sigma} b_{\mu} dx^{\mu} \ ,
\ee
where $\Sigma_i$ is a small region around the atom. This result 
contains an additional surface term when
compared with the existing literature \cite{GAS1,NEL5}. This surface
term will only vanish if defects successfully fully screen out the
Gaussian curvature.  As we will see in Eq.~\ref{ground_stat}, this  
is the condition that minimizes the energy of the model. So, at very
low temperatures, the surface term will be small.

\begin{figure}[htb]
\epsfxsize=2 in \centerline{\epsfbox{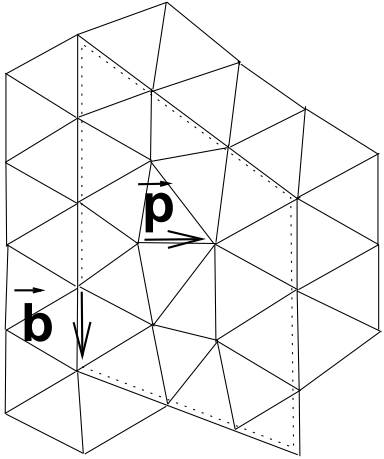}}
\caption{ A dislocation as a tightly bound disclination pair. The dotted
line is the Burgers circuit. It is apparent that the dipole moment 
${\bf p}$ is perpendicular to the Burgers vector ${\bf b}$. }
\label{fig__dislocation}
\end{figure}

Comparing Eq.~\ref{fancy_Gauss} with \cite{NEL5}, there is a somewhat 
disturbing factor $\frac{3}{\pi}$ in the last term. In a fine grained 
description of the model, one regards a
dislocation as a composite object, a tightly bound pair of opposite 
disclinations (see Fig.~\ref{fig__dislocation}).
Defining the dipole moment ${\bf p}$ as
\be\label{dipole_mom}
p^{\mu}=\frac{3}{\pi} {\epsilon^{\mu}}_{\nu} b^{\nu} \ ,
\ee
where $b^{\nu}$ is the microscopic Burgers vector, we can Taylor
expand the density of two free disclinations of opposite charge as
\be\label{disc_disl}
\frac{\pi}{3}
\left\{ \delta({\bf x}+{\bf p})-\delta({\bf x}) \right\} =
\epsilon^{\mu \nu} b_{\mu} \parp_{\nu} \delta({\bf x})+{\cal O}(a^2) \ ,
\ee
where we neglect higher order terms in the lattice spacing $a$. The
right hand side of this last equation corresponds exactly to an isolated dislocation. 
The factor $\frac{3}{\pi}$ is the slight displacement 
of the center of the disclinations necessary to render the right dislocation
density Eq.~\ref{disc_disl}. As a byproduct, we derive the
result that the dipole moment ${\bf p}$ is perpendicular to the
microscopic Burgers vector, a result which is also apparent from 
Fig.~\ref{fig__dislocation}.

For our purposes, however, it will be more convenient to use 
Eq.~\ref{Final_form} in combination with Eq.~\ref{actu_omega}, which
yields 
\be\label{good_rel}
b=-d \theta-\Omega_{sing}+\Omega^L \ .
\ee

This completes our geometrical development of the structure of
bond-orientational order. 

\section{The properties of the hexatic phase}\label{SECT__hex}

The previous section was somewhat formal. We derived some general
results valid in the presence of bond-orientational order. 
Further progress requires making additional physical assumptions 
about the interactions of dislocations and disclinations with the 
underlying geometry and topology.

For the simplest case of a flat monolayer 
a free energy was proposed in \cite{NEL1,YOU1} (see
\cite{NEL2} for a review). We now examine this proposal within the 
framework of the geometric formalism developed here. We use a completely
covariant language with the aid of the mathematical concepts sketched in
Sec.~\ref{sect__diff_geom}. Although this is not necessary for
the case of flat space, it is directly applicable to general
geometries.

\subsection{Debye-Huckel approximation on a flat monolayer}

On a flat monolayer, the metric of Eq.~\ref{met_surface} is the standard
Euclidean metric
\be\label{euclid_met}
ds^2=dx^2+dy^2 \ .
\ee
In this case $\vap^1=dx$ and $\vap^2=dy$. 

Let us first treat the simpler case of vanishing free disclination 
density. The bond angle as a function of the Burgers vector density is
found by inverting Eq.~\ref{good_rel}: 
\be\label{flat_dens}
\theta({\bf x})=\partial_{\mu}(\frac{1}{\Delta}b_{\mu})=
-\frac{1}{2\pi}\int d {\bf x'}
\frac{ {\bf b}({\bf x}')({\bf x-x'})}{|{\bf x - x'} |^2 } \ ,
\ee
where $\frac{1}{\Delta}=1/(2\pi)\log(|x-x'|)$. This result is
in agreement with the results in \cite{NEL1,YOU1,NEL2}. 
We emphasize here that this this result follows
from the Cartan structure equations alone, with no further physical 
assumptions. It provides a non-trivial consistency check for our
geometrical interpretation of the hexatic phase.

Additional constraints follow from Eq.~\ref{Gauss_form}. With no free 
disclinations it implies
\be\label{b_irrot}
db=0 \Longleftrightarrow \epsilon^{\mu \nu} \parp_{\mu} b_{\nu}=0 \ .
\ee
Geometry thus constrains the Burgers vector distribution 
${\bf b}({\bf x})$ to be irrotational. 

The free energy of the crystal, within a Debye-Huckel approximation,
was derived in \cite{NEL1,YOU1}: 
\be\label{stand_free}
{\cal H}/T=\frac{K_B}{2}(db|\frac{1}{\Delta^2} db)+\frac{E_ca^2}{T}
(b|b) \ ,
\ee
where the inner product between forms was defined in 
Eq.~\ref{inner_prod} and the Laplacian in Eq.~\ref{def_laplacian}.
$K_B$ is the 2-dimensional Young modulus,
$a$ is the original lattice spacing and $E_c$ the core energy of a
single dislocation.
The RG recursion relations of the 2d KTNHY melting transition 
\cite{NEL1} imply that in the hexatic phase 
\be\label{kb_infrared}
{\rm lim}_{{\bf p}\rightarrow 0} K_B({\bf p})=0 \ .
\ee
If now we consider the bond angle as the fundamental variable,
Eq.~\ref{good_rel} and Eq.~\ref{b_irrot} imply a simple $XY$ model
for the energy of the bond angle variable
\be\label{hex_xy}
{\cal H}/T=\frac{K_A}{2} \int \parp_{\mu} \theta \parp^{\mu} \theta \ ,
\ee
with hexatic stiffness $K_A=\frac{2 E_c a^2}{T}$ \cite{NEL1,YOU1}.

The irrotationality of ${\bf b}$ is a general constraint to be satisfied 
in any hexatic regime with vanishing free disclination density.
It has additional consequences. We already noted that
a dislocation may be regarded as a tightly bound disclination pair. 
The dipole vector ${\bf p}$, defined in Eq.~\ref{dipole_mom}, is
perpendicular to the Burgers vector ${\bf b}$. If we consider a
closed loop ${\cal C}$, we can compute the circulation for ${\bf p}$
\be\label{loop}
\int_{\cal C} {\bf p} d {\bf s}= \int_{\cal C} \epsilon^{\mu \nu} b_{\nu} dx_{\mu} =
\int_{\Sigma} dxdy \Delta \theta \ ,
\ee
where $\Sigma$ is the two dimensional region enclosed by the closed 
curve ${\cal C}$. The resultant integral is generally not zero. 
This means that dipoles form grain boundaries of 
consecutive fives and sevens. If there are no free disclinations 
these strings cannot self-intersect. 

In the presence of free disclinations, Eq.~\ref{b_irrot} gets modified
to
\be\label{b_now_rot}
db=s dx\wedge dy \Longleftrightarrow 
\epsilon^{\mu \nu} \parp_{\mu} b_{\nu}({\bf x})=s({\bf x}) \ .
\ee
The free energy for the bond angle variable now reads
\be\label{hex_discl}
{\cal H}/T=\frac{K_B}{2}(s|\frac{1}{\Delta}s)+
\frac{K_A}{2}(d\theta+\Omega_{sing}|d\theta+\Omega_{sing}) \ .
\ee
The first term leads to a strong binding of free disclinations, since it corresponds to
a $|{\bf x}|^2\ln|{\bf x}|$ disclination-disclination interaction . 
Recalling Eq.~\ref{kb_infrared}, however, we see that this term is
absent. The long-wavelength description of the hexatic phase is given
by the standard XY model \cite{NEL1}. 

The hexatic phase can be characterized by the correlator
\cite{HALP1}
\be\label{C_function}
C({\bf r})=\langle \sigma({\bf r}) \sigma({\bf 0}) \rangle \ ,
\ee
with
\be\label{top_dens}
\sigma({\bf r})=\sum_{l} {\rm Ind}_l \delta^2({\bf r}-{\bf R}_l)=
\frac{1}{2\pi} s({\bf x}) \ ,
\ee
where ${\rm Ind}_l$ is the index of the order parameter. The last identity
follows from the definition of the index of a vector field.
The order parameter Eq.~\ref{C_function} is thus the correlation function 
of the free disclination density, the hexatic curvature.
To lowest order in the fugacity, we have in the hexatic phase
\be\label{hexatic_cord}
\langle s({\bf x}) s({\bf 0}) \rangle \sim \frac{y^2}{|{\bf x}|^{\eta}}
\ ,
\ee
where $y$ is the fugacity of the vortices. In the isotropic liquid 
phase \cite{HALP1}
\be\label{isotropic_liq}
\langle s({\bf x}) s({\bf 0}) \rangle \sim e^{-\frac{|{\bf x}|}{\xi}} ,
\ee
where $\xi \sim e^{\frac{a}{T-T_c}}$ is the correlation length of the
model.

\subsection{Hexatic order in an arbitrary geometry}

In the previous section we recast standard results for a flat monolayer 
in the geometric framework developed in this paper. These results may now
be extended directly to the case of an arbitrary geometry.
The metric
\be\label{any_met}
ds^2=g_{\mu \nu} dx^{\mu} dx^{\nu} \ ,
\ee
takes a general form with metric coefficients given by
Eq.~\ref{met_surface}.

The first step is to find the free energy. This is easily obtained
from Eq.~\ref{stand_free} using the covariant definition of the 
operators involved, as described in sect.~\ref{sect__diff_geom}
We start by defining the form
\be\label{Gauss_discl}
\rho({\bf x})=\left\{ K({\bf x})-s({\bf x}) \right\} \Omega_M \ ,
\ee
expressing the difference between Gaussian curvature and free 
disclination density. In Eq.~\ref{stand_free} we deliberately wrote 
the free energy of a simple flat monolayer in terms of the torsion 
degrees of freedom. It is now extremely simple to generalize it to 
the case of a general geometry by making use the general relation 
Eq.~\ref{good_rel} that holds for an arbitrary geometry
to obtain
\be\label{hex_discl_geom}
{\cal H}/T=\frac{K_B}{8}(\rho|\frac{1}{\Delta^2} \rho)+
\frac{K_A}{2}(d\theta+\Omega_{sing}-\Omega^L|
d\theta+\Omega_{sing}-\Omega^L) \ ,
\ee
where $\Omega^L$ is the connection form of the Levi-Civita connection.
This corresponds to the free energy of a 2D crystal on an arbitrary
geometry. This result is identical to that obtained by integrating 
out the in-plane phonons \cite{NEL4} in the absence of free
disclinations, and generalizing to include these additional degrees of
freedom. The free energy Eq.~\ref{hex_discl_geom} 
corresponds interacting dislocations and disclinations  
in the crystalline phase. At zero temperature, one can substitute the
bond orientational order parameter $\theta$ by its minimum value
$\theta=\frac{1}{\Delta}\nabla_{\mu}\Omega^{\mu}$, and we get
\be\label{hex_discl_geom_t0}
{\cal H}/T=\frac{K_B}{8}(\rho|\frac{1}{\Delta^2} \rho)+
\frac{K_A}{2}(\rho|\frac{1}{\Delta} \rho) \ .
\ee
Free positive disclinations are attracted to positive curvature regions 
and negative disclinations to negative curvature regions. The ground state
is defined by the equation
\be\label{ground_stat}
\rho=0 \Rightarrow s({\bf x})=K({\bf x}) \ .
\ee
Defects will arrange themselves to optimally screen out 
the Gaussian curvature. 

As for the case of a flat monolayer (Eq.~\ref{kb_infrared}), the
hexatic phase requires that in the infrared limit
\be\label{kb_infrared_arb}
K_B=0 \ .
\ee
The free energy then becomes 
\bea\label{hexatic_energy}
{\cal H}/T&=&
\frac{K_A}{2}(d\theta+\Omega_{sing}-\Omega^L|
d\theta+\Omega_{sing}-\Omega^L)
\nonumber
\\
&=&\frac{K_A}{2}\int d {\bf x} \sqrt{g}
g^{\mu \nu}(\partial_{\mu} \theta+\Omega_{sing}-\Omega^L_{\mu})
(\partial_{\nu} \theta+\Omega_{sing}-\Omega^L_{\nu}) \ ,
\eea
which is the hexatic free energy first considered in \cite{NEL4}. 

It is beyond the scope of this paper to investigate the precise mechanism 
by which Eq.~\ref{kb_infrared_arb} is implemented in an arbitrary geometry,
but we show instead that it is a necessary condition for the hexatic
phase to be realized.

In analogy with the flat monolayer, one may compute the order parameter
Eq.~\ref{C_function} to lowest order in the fugacity,
\bea\label{arb_geom}
\langle s({\bf x}) s({\bf 0}) \rangle&=&-2\left(\frac{\pi}{3}\right)^2 y^2
e^{-\frac{K_A}{2}\sum_{{\bf x}_i \neq {\bf x}_j} \left(\frac{1}{\Delta}
\right)_{{\bf x_i} {\bf x_j}}} e^{K_A\sum_{{\bf i}=1}^2
\int d^2 {\bf x'} \frac{q_i}{\Delta}_{{\bf x} {\bf x'}}
\sqrt{g}K({\bf x})} \ ,
\nonumber\\
&&
\eea
which reduces to Eq.~\ref{hexatic_cord} in the case of a flat 
monolayer. 

It is worthwhile to recall at this point that 
if fluctuations in the geometry were allowed, we should include an
additional bending rigidity or extrinsic curvature term 
\be\label{extr_curv}
\frac{{\cal H}_{ex}}{T}=\frac{\kappa}{2}\int d {\bf x} 
\sqrt{g} {\vec H}^2 \ ,
\ee
where ${\vec H}$ is the Mean curvature of the surface. In this case
we average over all possible fluctuations in geometry. It seems that
Dislocations have a finite energy \cite{NEL3} and therefore they proliferate 
at any temperature, see also \cite{DAV1}.  

\section{Conclusions and Outlook}

A full understanding of defects in curved geometries is a challenging
subject with many novel features not encountered in the plane.      
In this paper, we presented a geometrical interpretation of
bond-orientational order leading to a variety of relations between the
underlying geometry, the topology and the nature of the defects. 
The formalism developed in this paper has already been applied very
successfully to the case of
defect arrays in spherical crystals \cite{BNT:99}. Further
applications are currently being explored. 
The most challenging open problem may be the generalization of the
KTNHY renormalization group flows to fixed or dynamical curved
geometries. The spherical crystal itself gives rise to a very rich
structure of defects in the ground state \cite{BNT:99} and other
geometries are very likely to lead to even more novel results.

\bigskip
\bigskip

{\bf Acknowledgements}
\medskip
\noindent 
The research of M.B. and A.T. was supported by the U.S. Department of
Energy under contract No. DE-FG02-85ER40237.

\newpage

\appendix

\section{The hexatic cone}

A sample in the hexatic phase on a template having the geometry 
of a cone is the simplest case where the present formalism 
may be applied. It provides an interesting benchmark as the cone has 
been the subject of much interest \cite{KAR,NEL7,LUB1}. 
For some range of parameters, an isolated disclination forces the 
flat monolayer to buckle; a positive disclination is optimally screened 
by the Gaussian curvature located at the cone. It is more convenient to 
work in polar coordinates $(r,\psi)$, the metric being
\be\label{cone_met}
ds^2=(1+m^2)dr^2+r^2d\psi^2 \ ,
\ee
where $m$ is related to the aperture angle of the cone $\zeta$ by 
$\tan \zeta=\frac{1}{m}$.

\begin{figure}[htb]
\epsfxsize=3 in \centerline{\epsfbox{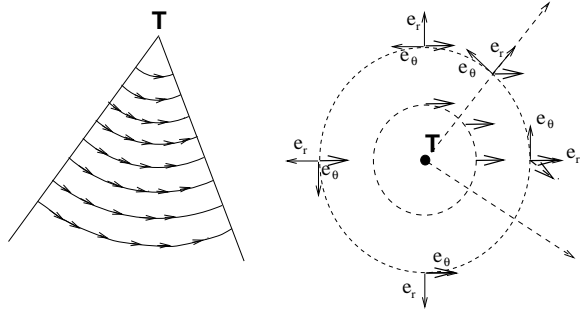}}
\caption{Distribution of torsion vectors in the cone. Plot of the order
parameter in polar coordinates $(r,\psi)$. The order parameter is the
same as in a plane despite the Gaussian curvature located at the tip $T$ of 
the cone.}
\label{fig__cone}
\end{figure}

One readily finds $\vap^1=\sqrt{1+m^2} dr$, $\vap^2=r d\psi$. The Levi-Civita
connection is $\Omega^L=-\frac{1}{\sqrt{1+m^2}}d\psi$. The actual hexatic 
connection is flat, corresponding to a plane
\be\label{hex_conn}
\Omega=-d\psi+d \phi \rightarrow \theta(r,\psi)=-\psi+\phi(r,\psi) \ ,
\ee
where $\phi$ is regular $\phi(r,\psi+2\pi)=\phi(r,\psi)$ and
corresponds to the different Burgers vectors distributions one may have 
in a cone. The lowest energy solution is $\phi=0$, and this is the only 
case we consider. The torsion vector distribution may be computed easily,
\be\label{cone_dist}
{\bf b}({\bf x})=(1-\frac{1}{\sqrt{1+m^2}})\frac{1}{r} e_{\psi} \ ,
\ee
Torsion vectors are tangent to circles centered at the tip of the cone,
as depicted in Fig.~\ref{fig__cone}. It is also illuminating to plot the order 
parameter, as illustrated in Fig.~\ref{fig__cone}. The parallel
transported bond angle is 
completely insensitive to the Gaussian curvature located
at the tip of the cone.
As a further cross check, we can compute the hexatic energy for 
this configuration, using the free energy Eq.~\ref{hexatic_energy}.
Assuming a lattice spacing $a$ as an ultraviolet cut-off and
the radius $R$ of the cone as an infrared cut-off, we get
\bea\label{cone_energy}
E_{hex}&=&\frac{K_A}{2}\int^{R}_a \frac{dr}{r}(1+m^2)^{\frac{1}{2}}
\int^{2 \pi}_0 d \psi(-1+\frac{1}{\sqrt{1+m^2}})^2
\nonumber\\
&=& \pi K_A(1-\frac{1}{\sqrt{1+m^2}})^2(1+m^2)^{1/2} \ln\frac{R}{a} \ .
\eea
This result has already been obtained by other authors 
\cite{KAR,NEL7,LUB1} by solving the equations of motion
for the order parameter and computing the Green's function.
Instead we have obtained it by constructing the order parameter by parallel
transport. As a byproduct, we are able to compute the torsion vector 
distribution in the cone.

If there is an isolated disclination located at the tip of the cone,
the hexatic connection is no longer flat, as there is hexatic
curvature located at the tip of the cone. Parallel transport is now
ambiguous if we encircle the tip of the cone. We get,
\be\label{hex_conn_discl}
\Omega=(\frac{q_i}{6}-1)d\psi \rightarrow 
\theta(r,\psi)=(\frac{q_i}{6}-1) \psi \ .
\ee
Following the same steps as before we obtain a free
energy of the form given in Eq.~\ref{cone_energy}.

\end{document}